\documentclass[reprint,aps,prl,twocolumn,superscriptaddress,showpacs,preprintnumbers,amsmath,amssymb,floatfix]{revtex4-1}
\usepackage{graphicx}  
\usepackage[mathlines]{lineno}
\usepackage{bm}        
\usepackage{amssymb}   
\usepackage{xcolor}
\usepackage{xspace} 
\usepackage{wasysym}
\usepackage[colorlinks,linkcolor=blue,citecolor=blue,urlcolor=blue]{hyper ref}

\begin{document}


\title{CDMSlite: A Search for Low-Mass WIMPs using Voltage-Assisted Calorimetric Ionization Detection in the SuperCDMS Experiment}

\affiliation{Division of Physics, Mathematics, \& Astronomy, California Institute of Technology, Pasadena, CA 91125, USA} 
\affiliation{Fermi National Accelerator Laboratory, Batavia, IL 60510, USA}
\affiliation{Lawrence Berkeley National Laboratory, Berkeley, CA 94720, USA}
\affiliation{Department of Physics, Massachusetts Institute of Technology, Cambridge, MA 02139, USA}
\affiliation{Pacific Northwest National Laboratory, Richland, WA 99352, USA}
 \affiliation{Department of Physics, Queen's University, Kingston ON, Canada K7L 3N6}
\affiliation{Department of Physics, Santa Clara University, Santa Clara, CA 95053, USA}
\affiliation{SLAC National Accelerator Laboratory/Kavli Institute for Particle Astrophysics and Cosmology, 2575 Sand Hill Road, Menlo Park 94025, CA}
\affiliation{Department of Physics, Southern Methodist University, Dallas, TX 75275, USA}
\affiliation{Department of Physics, Stanford University, Stanford, CA 94305, USA}
\affiliation{Department of Physics, Syracuse University, Syracuse, NY 13244, USA}
\affiliation{Department of Physics, Texas A\&M University, College Station, TX 77843, USA}
\affiliation{Departamento de F\'{\i}sica Te\'orica and Instituto de F\'{\i}sica Te\'orica UAM/CSIC, Universidad Aut\'onoma de Madrid, 28049 Madrid, Spain}
\affiliation{Department of Physics, University of California, Berkeley, CA 94720, USA}
\affiliation{Department of Physics, University of California, Santa Barbara, CA 93106, USA}
\affiliation{Department of Physics, University of Colorado Denver, Denver, CO 80217, USA}
\affiliation{Department of Physics, University of Evansville, Evansville, IN 47722, USA}
\affiliation{Department of Physics, University of Florida, Gainesville, FL 32611, USA}
\affiliation{Department of Physics, University of Illinois at Urbana-Champaign, Urbana, IL 61801, USA}
\affiliation{School of Physics \& Astronomy, University of Minnesota, Minneapolis, MN 55455, USA}
 \affiliation{Department of Physics, University of South Dakota, Vermillion, SD 57069, USA}

\author{R.~Agnese} \affiliation{Department of Physics, University of Florida, Gainesville, FL 32611, USA}
\author{A.J.~Anderson} \affiliation{Department of Physics, Massachusetts Institute of Technology, Cambridge, MA 02139, USA}
\author{M.~Asai} \affiliation{SLAC National Accelerator Laboratory/Kavli Institute for Particle Astrophysics and Cosmology, 2575 Sand Hill Road, Menlo Park 94025, CA}
\author{D.~Balakishiyeva} \affiliation{Department of Physics, University of Florida, Gainesville, FL 32611, USA}
\author{R.~Basu~Thakur~} \email{Corresponding author: ritoban@fnal.gov}\affiliation{Fermi National Accelerator Laboratory, Batavia, IL 60510, USA}\affiliation{Department of Physics, University of Illinois at Urbana-Champaign, Urbana, IL 61801, USA}
\author{D.A.~Bauer} \affiliation{Fermi National Accelerator Laboratory, Batavia, IL 60510, USA}
\author{J.~Billard} \affiliation{Department of Physics, Massachusetts Institute of Technology, Cambridge, MA 02139, USA}
\author{A.~Borgland} \affiliation{SLAC National Accelerator Laboratory/Kavli Institute for Particle Astrophysics and Cosmology, 2575 Sand Hill Road, Menlo Park 94025, CA}
\author{M.A.~Bowles} \affiliation{Department of Physics, Syracuse University, Syracuse, NY 13244, USA}
\author{D.~Brandt} \affiliation{SLAC National Accelerator Laboratory/Kavli Institute for Particle Astrophysics and Cosmology, 2575 Sand Hill Road, Menlo Park 94025, CA}
\author{P.L.~Brink} \affiliation{SLAC National Accelerator Laboratory/Kavli Institute for Particle Astrophysics and Cosmology, 2575 Sand Hill Road, Menlo Park 94025, CA}
\author{R.~Bunker} \affiliation{Department of Physics, Syracuse University, Syracuse, NY 13244, USA}
\author{B.~Cabrera} \affiliation{Department of Physics, Stanford University, Stanford, CA 94305, USA}
\author{D.O.~Caldwell} \affiliation{Department of Physics, University of California, Santa Barbara, CA 93106, USA}
\author{D.G.~Cerdeno} \affiliation{Departamento de F\'{\i}sica Te\'orica and Instituto de F\'{\i}sica Te\'orica UAM/CSIC, Universidad Aut\'onoma de Madrid, 28049 Madrid, Spain}

\author{H.~Chagani} \affiliation{School of Physics \& Astronomy, University of Minnesota, Minneapolis, MN 55455, USA}
\author{J.~Cooley} \affiliation{Department of Physics, Southern Methodist University, Dallas, TX 75275, USA}
\author{B.~Cornell} \affiliation{Division of Physics, Mathematics, \& Astronomy, California Institute of Technology, Pasadena, CA 91125, USA}
\author{C.H.~Crewdson} \affiliation{Department of Physics, Queen's University, Kingston ON, Canada K7L 3N6}
\author{P.~Cushman} \affiliation{School of Physics \& Astronomy, University of Minnesota, Minneapolis, MN 55455, USA}
\author{M.~Daal} \affiliation{Department of Physics, University of California, Berkeley, CA 94720, USA}
\author{P.C.F.~Di~Stefano} \affiliation{Department of Physics, Queen's University, Kingston ON, Canada K7L 3N6}
\author{T.~Doughty} \affiliation{Department of Physics, University of California, Berkeley, CA 94720, USA}
\author{L.~Esteban} \affiliation{Departamento de F\'{\i}sica Te\'orica and Instituto de F\'{\i}sica Te\'orica UAM/CSIC, Universidad Aut\'onoma de Madrid, 28049 Madrid, Spain}

\author{S.~Fallows} \affiliation{School of Physics \& Astronomy, University of Minnesota, Minneapolis, MN 55455, USA}
\author{E.~Figueroa-Feliciano} \affiliation{Department of Physics, Massachusetts Institute of Technology, Cambridge, MA 02139, USA}
\author{G.L.~Godfrey} \affiliation{SLAC National Accelerator Laboratory/Kavli Institute for Particle Astrophysics and Cosmology, 2575 Sand Hill Road, Menlo Park 94025, CA}
\author{S.R.~Golwala} \affiliation{Division of Physics, Mathematics, \& Astronomy, California Institute of Technology, Pasadena, CA 91125, USA}
\author{J.~Hall} \affiliation{Pacific Northwest National Laboratory, Richland, WA 99352, USA}
\author{H.R.~Harris} \affiliation{Department of Physics, Texas A\&M University, College Station, TX 77843, USA}
\author{S.A.~Hertel} \affiliation{Department of Physics, Massachusetts Institute of Technology, Cambridge, MA 02139, USA}
\author{T.~Hofer} \affiliation{School of Physics \& Astronomy, University of Minnesota, Minneapolis, MN 55455, USA}
\author{D.~Holmgren} \affiliation{Fermi National Accelerator Laboratory, Batavia, IL 60510, USA}
\author{L.~Hsu} \affiliation{Fermi National Accelerator Laboratory, Batavia, IL 60510, USA}
\author{M.E.~Huber} \affiliation{Department of Physics, University of Colorado Denver, Denver, CO 80217, USA}
\author{A.~Jastram} \affiliation{Department of Physics, Texas A\&M University, College Station, TX 77843, USA}
\author{O.~Kamaev} \affiliation{Department of Physics, Queen's University, Kingston ON, Canada K7L 3N6}
\author{B.~Kara} \affiliation{Department of Physics, Southern Methodist University, Dallas, TX 75275, USA}
\author{M.H.~Kelsey} \affiliation{SLAC National Accelerator Laboratory/Kavli Institute for Particle Astrophysics and Cosmology, 2575 Sand Hill Road, Menlo Park 94025, CA}
\author{A.~Kennedy} \affiliation{School of Physics \& Astronomy, University of Minnesota, Minneapolis, MN 55455, USA}
\author{M.~Kiveni} \affiliation{Department of Physics, Syracuse University, Syracuse, NY 13244, USA}
\author{K.~Koch} \affiliation{School of Physics \& Astronomy, University of Minnesota, Minneapolis, MN 55455, USA}
\author{B.~Loer} \affiliation{Fermi National Accelerator Laboratory, Batavia, IL 60510, USA}
\author{E.~Lopez~Asamar} \affiliation{Departamento de F\'{\i}sica Te\'orica and Instituto de F\'{\i}sica Te\'orica UAM/CSIC, Universidad Aut\'onoma de Madrid, 28049 Madrid, Spain}

\author{R.~Mahapatra} \affiliation{Department of Physics, Texas A\&M University, College Station, TX 77843, USA}
\author{V.~Mandic} \affiliation{School of Physics \& Astronomy, University of Minnesota, Minneapolis, MN 55455, USA}
\author{C.~Martinez} \affiliation{Department of Physics, Queen's University, Kingston ON, Canada K7L 3N6}
\author{K.A.~McCarthy} \affiliation{Department of Physics, Massachusetts Institute of Technology, Cambridge, MA 02139, USA}
\author{N.~Mirabolfathi} \affiliation{Department of Physics, University of California, Berkeley, CA 94720, USA}
\author{R.A.~Moffatt} \affiliation{Department of Physics, Stanford University, Stanford, CA 94305, USA}
\author{D.C.~Moore} \affiliation{Division of Physics, Mathematics, \& Astronomy, California Institute of Technology, Pasadena, CA 91125, USA}
\author{P.~Nadeau} \affiliation{Department of Physics, Queen's University, Kingston ON, Canada K7L 3N6}
\author{R.H.~Nelson} \affiliation{Division of Physics, Mathematics, \& Astronomy, California Institute of Technology, Pasadena, CA 91125, USA}
\author{K.~Page} \affiliation{Department of Physics, Queen's University, Kingston ON, Canada K7L 3N6}
\author{R.~Partridge} \affiliation{SLAC National Accelerator Laboratory/Kavli Institute for Particle Astrophysics and Cosmology, 2575 Sand Hill Road, Menlo Park 94025, CA}
\author{M.~Pepin} \affiliation{School of Physics \& Astronomy, University of Minnesota, Minneapolis, MN 55455, USA}
\author{A.~Phipps} \affiliation{Department of Physics, University of California, Berkeley, CA 94720, USA}
\author{K.~Prasad} \affiliation{Department of Physics, Texas A\&M University, College Station, TX 77843, USA}
\author{M.~Pyle} \affiliation{Department of Physics, University of California, Berkeley, CA 94720, USA}
\author{H.~Qiu} \affiliation{Department of Physics, Southern Methodist University, Dallas, TX 75275, USA}
\author{W.~Rau} \affiliation{Department of Physics, Queen's University, Kingston ON, Canada K7L 3N6}
\author{P.~Redl} \affiliation{Department of Physics, Stanford University, Stanford, CA 94305, USA}
\author{A.~Reisetter} \affiliation{Department of Physics, University of Evansville, Evansville, IN 47722, USA}
\author{Y.~Ricci} \affiliation{Department of Physics, Queen's University, Kingston ON, Canada K7L 3N6}
\author{T.~Saab} \affiliation{Department of Physics, University of Florida, Gainesville, FL 32611, USA}
\author{B.~Sadoulet} \affiliation{Department of Physics, University of California, Berkeley, CA 94720, USA}\affiliation{Lawrence Berkeley National Laboratory, Berkeley, CA 94720, USA}
\author{J.~Sander} \affiliation{Department of Physics, University of South Dakota, Vermillion, SD 57069, USA}
\author{K.~Schneck} \affiliation{SLAC National Accelerator Laboratory/Kavli Institute for Particle Astrophysics and Cosmology, 2575 Sand Hill Road, Menlo Park 94025, CA}
\author{R.W.~Schnee} \affiliation{Department of Physics, Syracuse University, Syracuse, NY 13244, USA}
\author{S.~Scorza} \affiliation{Department of Physics, Southern Methodist University, Dallas, TX 75275, USA}
\author{B.~Serfass} \affiliation{Department of Physics, University of California, Berkeley, CA 94720, USA}
\author{B.~Shank} \affiliation{Department of Physics, Stanford University, Stanford, CA 94305, USA}
\author{D.~Speller} \affiliation{Department of Physics, University of California, Berkeley, CA 94720, USA}
\author{A.N.~Villano} \affiliation{School of Physics \& Astronomy, University of Minnesota, Minneapolis, MN 55455, USA}
\author{B.~Welliver} \affiliation{Department of Physics, University of Florida, Gainesville, FL 32611, USA}
\author{D.H.~Wright} \affiliation{SLAC National Accelerator Laboratory/Kavli Institute for Particle Astrophysics and Cosmology, 2575 Sand Hill Road, Menlo Park 94025, CA}
\author{S.~Yellin} \affiliation{Department of Physics, Stanford University, Stanford, CA 94305, USA}
\author{J.J.~Yen} \affiliation{Department of Physics, Stanford University, Stanford, CA 94305, USA}
\author{B.A.~Young} \affiliation{Department of Physics, Santa Clara University, Santa Clara, CA 95053, USA}
\author{J.~Zhang} \affiliation{School of Physics \& Astronomy, University of Minnesota, Minneapolis, MN 55455, USA}

\collaboration{The SuperCDMS Collaboration} 
\noaffiliation

\begin{abstract}
SuperCDMS is an experiment designed to directly detect Weakly Interacting Massive Particles (WIMPs), a favored candidate for dark matter ubiquitous in the Universe. In this paper, we present WIMP-search results using a calorimetric technique we call CDMSlite, which relies on voltage-assisted Luke-Neganov amplification of the ionization energy deposited by particle interactions. The data were collected with a single 0.6 kg germanium detector running for 10 live days at the Soudan Underground Laboratory.  A low energy threshold of $170\;\text{eV}_{\text{ee}}$  (electron equivalent) was obtained, which allows us to constrain new WIMP-nucleon spin-independent parameter space for WIMP masses below 6 GeV/$c^2$.
\end{abstract}

\maketitle

Independent astrophysical surveys and cosmological studies confirm that dark matter constitutes $27\%$ of the energy density of the Universe (reviewed in~\cite{Pdg2012}). Weakly Interacting Massive Particles (WIMPs) are one of the favored particle candidates for dark matter. 
Theoretical predictions for WIMP masses, and for WIMP-interaction cross sections on normal matter, both span many orders of magnitude.
%
%
However, WIMPs may elastically scatter off nuclei with enough energy, and at a sufficient rate, to be detected by laboratory detectors~\cite{PhysRevD.31.3059}.  Measurements  of the nuclear-recoil energy spectrum by these experiments can constrain the properties of WIMP dark matter ~\cite{ Lewin199687, Jungman1996195, 2012particle}. 

Some extensions to the Standard Model of particle physics predict new stable particles, that could have been produced in the early Universe, with the properties needed to explain the current dark matter density~\cite{Jungman1996195}. The DAMA~\cite{Bernabei:2010uq,Savage:2008er}, CoGeNT~\cite{PhysRevD.88.012002}, CRESST II~\cite{Angloher:2012fk} and CDMS II Si~\cite{Agnese:2013rvf} experiments have reported excesses of events at low energies compared with their background models, hinting at the possible existence of low-mass WIMPs.  The diffuse gamma-ray emission from the galactic center has also been interpreted as evidence for annihilation of light WIMPs~\cite{2013arXiv1302.6589H}. There have been several attempts to reconcile these hints with a low-mass WIMP hypothesis ~\cite{PhysRevD.82.123509, PhysRevD.71.123520, Savage:2008er, PhysRevD.83.083517, 2011ApJ74195}, and many extensions to the Standard Model naturally prefer $\mathcal{O}(1-10) \;\text{GeV}/c^2$ dark matter~\cite{Nussinov198555,*PhysRevLett.68.741,*PhysRevD.79.115016,*Falkowski:2011fk,*doi:10.1142/S0217751X13300287,*Zurek:2013wia,Burgess2001709,*1475-7516-2008-10-034,*1126-6708-2009-05-036,doi:10.1142/S0218271804006449,*PhysRevD.78.043529,Holdom1986196,*PhysRevD.76.083519,*1126-6708-2008-12-104,*PhysRevD.77.087302,*Pospelov200853,*PhysRevD.79.115002,*PhysRevD.79.015014,*PhysRevD.80.035008,*Pospelov2009391,*Essig:2010ye,*PhysRevD.86.056009,Feng200837,*PhysRevLett.101.231301}.  


Direct detection of low-mass WIMPs is an experimental challenge requiring sensitivity to nuclear-recoil energies $\lesssim$1~keV.  For some technologies, such small energy depositions are indistinguishable from electronic noise.  Those with sufficient signal-to-noise are often limited by backgrounds whose intrinsic rates increase at low energies.  Further, the performance of background-discrimination techniques tends to degrade at energies near the electronic-noise level because of resolution smearing.  For WIMPs lighter than $\sim10\;\text{GeV}/c^2$, there are also nontrivial systematic uncertainties associated with detector response~\cite{PhysRevC.84.045805, Mei} and the galactic halo model~\cite{2013arXiv1304.6401M}.

The SuperCDMS experiment~\cite{Sander:2012nia} is located in the Soudan Underground Laboratory (rock overburden equivalent to 2090~m of water) and utilizes the CDMS~II experiment's infrastructure~\cite{soudan_setup}. SuperCDMS consists of fifteen 0.6~kg germanium ``iZIP'' detectors~\cite{izippaper, iZIP_rej, mpylephd}, arranged in five towers of three detectors each. Phonon and charge sensors are interleaved on both faces of the cylindrical crystals. The total phonon energy deposited in the crystals is measured by Transition Edge Sensors (TESs) connected to aluminum collection fins and read out by Superconducting Quantum Interference Devices (SQUIDs). During normal operation, we trigger on phonon signals $\gtrsim2$~keV with WIMP sensitivity optimized for the range 10~GeV/$c^2$--10~TeV/$c^2$.

The data described here were collected using a single iZIP detector operated at $56 (\pm 4)$~mK, in a new mode (CDMSlite, for CDMS Low Ionization Threshold Experiment) that yields significantly better sensitivity to WIMPs of mass $<$\,10\,GeV/$c^2$. This mode of operation uses a relatively high bias voltage across the detector, leading to a large Luke-Neganov~\cite{luke,neganov,wang:094504} amplification of the phonon signal~\cite{Akerib, isaila, Spooner1992382}. Any interaction depositing energy above the 0.6 eV germanium bandgap promotes electron-hole pairs to the conduction band. The number of pairs ($N_\text{eh}$) depends on the energy and type of recoiling particle. These charge carriers are collected at the two detector surfaces by applying a bias voltage ($V_b$). The work done in drifting the charge carriers, $N_\text{eh}eV_b$, is emitted as Luke-Neganov phonons ~\cite{luke,neganov, wang:094504}. Assuming all charges recombine at the electrodes, the total phonon energy collected for a given event is a sum of the energy from primary-recoil and recombination phonons ($E_r$) and from the Luke-Neganov phonons, 
\begin{equation} E_T=E_r+N_\text{eh}eV_b \label{eq:ln1}.
\end{equation}
 For electron recoils in Ge, the average excitation energy per charge pair is $\varepsilon_{\gamma}= 3 \; \text{eV}$. If the phonon energy is calibrated with respect to electron recoils, then it is labeled in electron-equivalent units, $\text{eV}_{\text{ee}}$ or keV$_\text{ee}$. 
 
 Normal operation of the iZIP detectors provides excellent event-by-event discrimination against electron-recoil backgrounds~\cite{iZIP_rej}, but with relatively high energy thresholds. The CDMSlite operating mode gives a substantial reduction in energy threshold and improvement in energy resolution, by using the phonon instrumentation to measure ionization. However, discrimination between nuclear and electron recoils via the simultaneous measurement of phonon and ionization signals was not possible because of the electric-field geometry used for this first CDMSlite data set.
 
The single detector used for this initial CDMSlite result was selected because of its good electronic noise resolution and low leakage current through the crystal. The noise was observed to increase slightly starting at $V_b \gtrsim 60$~V, and more rapidly for $V_b \gtrsim 85$~V. The operating bias, $V_b=69~$V, was chosen to optimize signal-to-noise. The total phonon energy for electron recoils is
  \begin{equation} E_T=E_r\times \left(1+\frac{eV_b}{\varepsilon_{\gamma}}\right)  \label{eq:ln2}. 
\end{equation}
For $V_b=69$~V, $E_T=E_r\times 24$, resulting in a baseline resolution $\sigma=14$~eV$_{\text{ee}}$.

The standard SuperCDMS electronics were not designed for bias voltages larger than 10~V.  For CDMSlite, custom electronics were implemented that held an entire detector face at the desired bias voltage. The other face was kept at ground potential and operated with the standard SuperCDMS electronics to measure the total phonon energy. The current hardware cannot read out the biased face, but both faces are instrumented with phonon absorbers. Thus, the phonon collection efficiency was only half of the collection efficiency in standard iZIP operation.
      
The CDMSlite detector was operated for a total of $15.7$~live days of WIMP search, with $^{133}$Ba gamma calibration data interspersed throughout. Additionally, the detector was twice exposed to neutrons from a $^{252}$Cf neutron source, resulting in sufficient activation ($^{70}$Ge $+$ n $\rightarrow$ $^{71}$Ge) to determine the energy scale and monitor stability. $^{71}$Ge primarily decays via K- and L-shell electron captures, yielding 10.36 and 1.29 $\text{keV}_{\text{ee}}$ cascades of x-rays and Auger electrons with total energy equal to the binding energy of the respective Ga electron shell. The measured gain matched the expectation for electron recoils with a $\times 24$ amplification reduced by readout of only one side of a two-sided detector. 
There was an $\sim 8$\% variation over time, which is believed to be due to humidity-dependent leakage currents in the CDMSlite electronics (an offline test confirmed that changing humidity leads to significant changes in leakage currents).
%
%
The 10.36~$\text{keV}_{\text{ee}}$ line was used to correct for the gain variation and to set the overall energy scale. Time periods when this line was not intense enough to monitor the gain, because of the length of time since the last neutron activation, were removed from this analysis, removing $3.2$~days of live time.  Immediately after biasing the detector, exponentially decaying leakage currents were observed, with time constants that varied from a few minutes to tens of minutes. Time periods up of to four of these time constants were excluded,  costing $2.2$~days of live time. After applying these data-selection criteria,  the remaining WIMP-search exposure was 10.3~live days.

A number of event-selection criteria were applied to these data. Events with time-coincident signals in the muon veto detectors were removed in this analysis. Multiple-scatter events, for which at least one other SuperCDMS detector had reconstructed energy more than 3~$\sigma$ above noise, were rejected. Electronic glitches, the majority of which cause multiple detectors to trigger, were removed. A class of small electronic glitches that triggered only single detectors was observed. These glitch pulses are sharper than phonon pulses originating from particle interactions in the detector, so events matching a glitch pulse-shape template were also rejected. Events in which low-frequency noise triggered were removed by requiring the pulse rise time to be consistent with those measured during calibration with ionizing radiation. The combined WIMP detection efficiency for these criteria, calculated from pulse-shape Monte Carlo simulations, $^{133}$Ba calibration data, and randomly triggered events spread uniformly throughout the physics run, is $ 98.5\%$ for phonon pulses above $110~\text{eV}_{\text{ee}}$.

\begin{figure}[t!!]
	\begin{center}
		\includegraphics[width=250 pt, trim =5 0 0 5, clip=true]{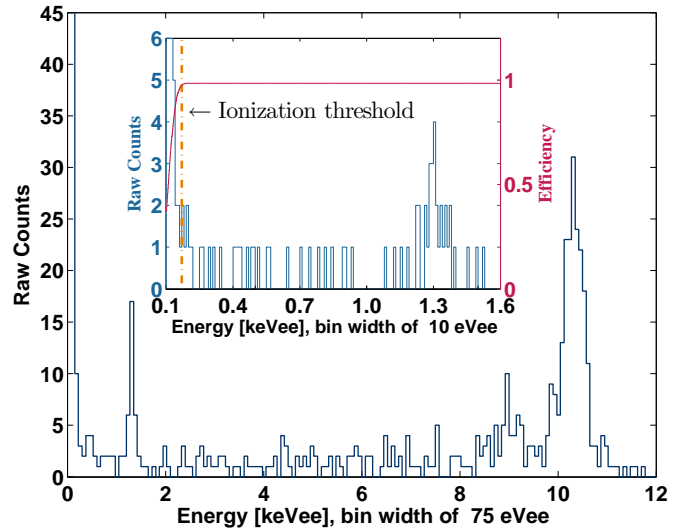}
				\caption{Recoil energy spectrum of WIMP-search events, after application of event-selection cuts. Inset: Low-energy spectrum in terms of raw counts (blue); also shown is the analysis efficiency. Both are expressed in $\text{keV}_{\text{ee}}$. The analysis threshold of $170\; \text{eV}_{\text{ee}}$ is indicated by the vertical dot-dashed line.  The resolution of the 1.3 keV line is 43~eV$_\text{ee}$ ($1\sigma$).} 
		\label{rawspect1}
	\end{center}
\end{figure}

The trigger efficiency was measured using low-energy events that passed these event-selection cuts. The efficiency was calculated with $^{133}$Ba calibration events triggered by another detector and verified with similar events from the WIMP-search data.  Because of the larger available counts, the calibration data were used to derive the final trigger efficiency. In this measurement, $50\%$ efficiency was reached at $108$ $\text{eV}_{\text{ee}}$. Low-frequency noise dominated the trigger rate below $ \sim100\; \text{eV}_{\text{ee}}$, well above the $14\; \text{eV}_{\text{ee}}$ baseline resolution limit.  The analysis threshold was set to $170\; \text{eV}_{\text{ee}}$, and the trigger efficiency is $100\%$ at, and above, this energy.  Figure~\ref{rawspect1} displays the measured spectrum up to 12~$\text{keV}_{\text{ee}}$. The inset in Fig.~\ref{rawspect1} shows the combined veto, event-selection, and trigger efficiencies, with the energy spectrum of WIMP-search events from 0.1 to 1.6~$\text{keV}_{\text{ee}}$.
The spectrum shows two main activation lines at 1.29 and 10.36~$\text{keV}_{\text{ee}}$, along with lines corresponding to cosmogenic activation: $8.98 \;\text{keV}_{\text{ee}}$ ($^{68}\text{Ga}$) and $9.66 \; \text{keV}_{\text{ee}}\;(^{68}\text{Zn})$. No other significant lines were found~\cite{activationlines}. Furthermore, the rate under 1~$\text{keV}_{\text{ee}}$ did not increase significantly after neutron calibration. The spectrum is relatively flat at low energies; however the average level is different above and below the 1.29~$\text{keV}_{\text{ee}}$ line. The average rate is 5.2~$\pm$~1~counts/$\text{keV}_{\text{ee}}$/kg-day between 0.2 and 1~$\text{keV}_{\text{ee}}$, and 2.9~$\pm$~0.3~counts/$\text{keV}_{\text{ee}}$/kg-day between 2 and 7~$\text{keV}_{\text{ee}}$. Further precise statements about the energy spectrum are limited by the low number of counts in the data presented here.

\begin{figure}[t!!]
	\begin{center}
		\includegraphics[width=248 pt, trim =5 0 0 8, clip=true]{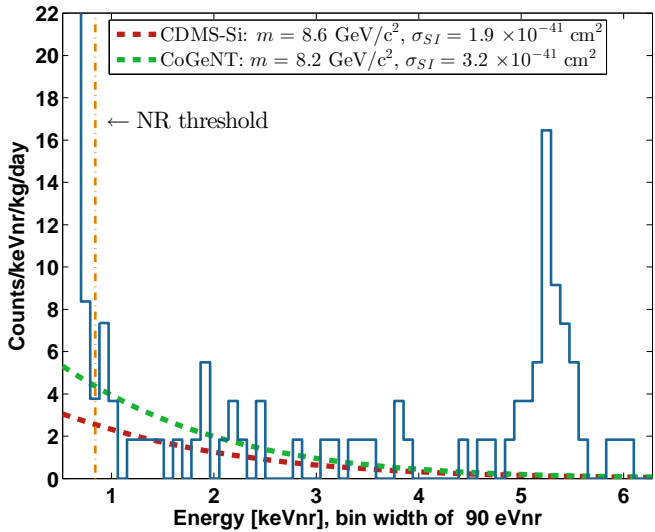}
		\caption{The efficiency-corrected WIMP-search energy spectrum is shown in $\text{keV}_{\text{nr}}$, and compared with expected rates for WIMPs with the most likely masses and cross sections suggested by the analysis of CoGeNT~\cite{PhysRevD.88.012002} and CDMS II Si~\cite{Agnese:2013rvf} data (dashed curves). Note that the $k=0.157$ Lindhard yield model was used to convert from an electron-equivalent to a nuclear-recoil-equivalent energy scale. The $170\; \text{eV}_{\text{ee}}$ ionization threshold translates to $841\; \text{eV}_{\text{nr}}$ (amber dot-dashed line). The $1.3~\text{keV}_{\text{ee}}$ activation line appears at $\sim5.3~\text{keV}_\text{nr}$.}
		\label{varwimprates1}
	\end{center}
\end{figure}

To use the energy spectrum shown in Fig.~\ref{rawspect1} to search for WIMPs, it must be converted to a nuclear-recoil-equivalent energy scale, with units denoted as $\text{keV}_{\text{nr}}$. We do so assuming 100\% charge collection for every event. The number of charges created by nuclear recoils is smaller than that for equivalent-energy electron recoils. This ``quenching'' can be parametrized as a reduction in the number of charges produced as $N_\text{eh}=E_\text{nr}Y(E_\text{nr})/\varepsilon_\gamma$, where $Y$ is the ionization yield, which measures the ionization energy per recoil energy, and is defined to be unity for electron recoils. The phonon energy can be converted to a nuclear-recoil-equivalent energy scale ($E_{\mathrm{nr}}$) using the equation
 \begin{equation}
E_\text{nr}=  E_\text{ee} \frac{(1 + \frac{eV_\text{b}}{\varepsilon_{\gamma}})}{(1 + \frac{eV_\text{b}}{\varepsilon_{\gamma}} Y(E_\text{nr}))}
\end{equation}
The ionization yield is not measured in this experiment, so a theoretical model is used. The most commonly used yield model is from Lindhard~\cite{Lindhard,Lewin199687},  given by the following formula for a nucleus with $Z$ protons and with atomic mass $A$:
\begin{equation}
Y(E_\text{nr} \text{(keV)}) = k\frac{g(\varepsilon)}{1+kg(\varepsilon)},
\end{equation}
with  $g(\varepsilon) = 3\varepsilon^{0.15} + 0.7\varepsilon^{0.6} + \varepsilon$, $\varepsilon = 11.5E_\text{nr}(\text{keV})Z^{-7/3}$ and $k = 0.133Z^{2/3}A^{-1/2}$. This gives $k=0.157$ for a germanium target. 
The constant $k$ is sometimes adjusted by experimenters to fit measurements.  Though other yield models, including simple power-law fits to data, have been used elsewhere~\cite{PhysRevD.86.051701,PhysRevD.88.012002}, we have carried out our conversion to nuclear-recoil equivalent using the standard Lindhard model, as recommended by Barker and Mei~\cite{Mei}. 
Under this assumption, the threshold is $841$~eVnr, with less than a $1.5$\% change from the $\sim 8$\% gain drift.
The resulting spectrum is shown in Fig.~\ref{varwimprates1} with examples of expected rates from two WIMP models.
\\
The region of interest used for limiting possible signal events from light WIMP scatters is between the 170~$\text{eV}_{\text{ee}}$ analysis threshold and 7~$\text{keV}_{\text{ee}}$. A  90\% C.L. upper limit on the spin-independent WIMP-nucleon cross section as a function of WIMP mass is calculated using the ``optimum interval" method~\cite{upper,*2007arXiv0709.2701Y},  using standard assumptions of a WIMP mass density of 0.3\, GeV/$c^2$/cm$^3$, a most probable WIMP velocity with respect to the galaxy of 220 km/s, a mean circular velocity of the Earth with respect to the galactic center of 232 km/s, a galactic escape velocity of 544 km/s, and the Helm form factor~\cite{Lewin199687}.
\begin{figure}[t!!]
	\begin{center}
		\includegraphics[width=250 pt, trim =0 0 0 0, clip=true]{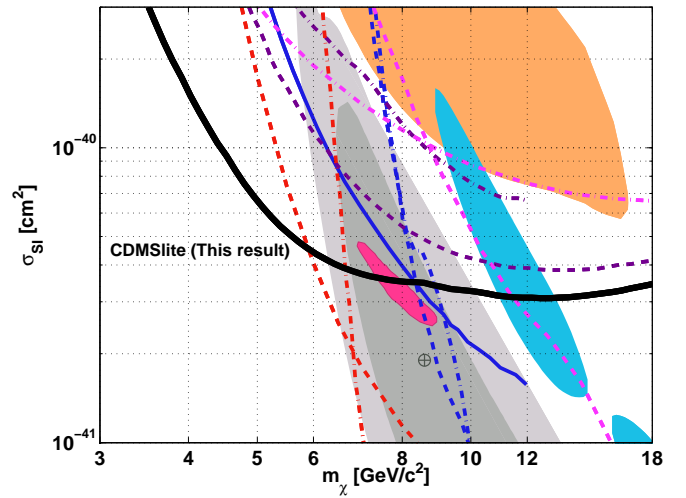}
		\caption{The 90\% upper confidence limit from the data presented here are shown with exclusion limits from other experiments. These are grouped as
		 Ge bolometers in blue:
		CDMS II Ge regular (dot-dash)~\cite{CDMSScience:2010},
		CDMS II Ge low threshold (solid)~\cite{PhysRevLett.106.131302},
		EDELWEISS II low threshold (dash)~\cite{PhysRevD.86.051701};
                    point-contact Ge detectors  in purple:
                   TEXONO (dash)~\cite{PhysRevLett.110.261301},
                   CDEX (dot-dash)~\cite{Zhao:2013xsf};
		 liquid Xenon in red:
		XENON100 (dot-dash)~\cite{PhysRevLett.109.181301},
		XENON10 S2 only (dash)~\cite{PhysRevLett.107.051301,*PhysRevLett.110.249901},
		LUX (solid)~\cite{Akerib:2013tjd};
		and other technologies in magenta:
		Low threshold reanalysis of CRESST II data (dot-dash)~\cite{PhysRevD.85.021301},
		PICASSO (dash)~\cite{Archambault2012153}.		
		The contours are from
		CDMS II Si (light and dark gray correspond to 68\% and 90\% CL regions respectively)~\cite{Agnese:2013rvf},
		CRESST II (blue)~\cite{Angloher:2012fk},
		DAMA (orange)~\cite{Bernabei:2010uq,Savage:2008er}, 
		CoGeNT (pink)~\cite{PhysRevD.88.012002}.}
		\label{offlimit_rito}
	\end{center}
\end{figure}

As shown in Fig.~\ref{offlimit_rito}, this analysis limits new WIMP parameter space for WIMP masses $< 6\; \text{GeV}/\text{c}^2$ and rules out portions of both the CDMS II Si~\cite{Agnese:2013rvf} and CoGeNT~\cite{PhysRevD.88.012002} contours. The CDMS II Si results had 3 WIMP candidate events in $\sim$140~kg-days, with an expected background of $\sim 0.5$ events. CoGeNT had an exposure of $\sim$~269 kg-days and performed a background subtraction for their results. These CDMSlite limits were obtained with a small net exposure of  $\sim6$~kg-days, minimal efficiency corrections, and no background subtraction.

\begin{figure}[t!]
	\begin{center}
		\includegraphics[width=250 pt, trim =0 0 0 11, clip=true]{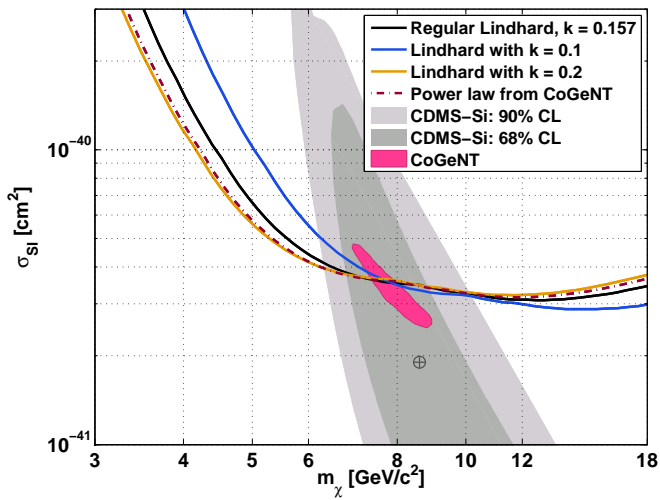}
		\caption{The effect of the choice of the yield model on the 90\% confidence level upper limit is shown. For the Lindhard model, the $0.170~\text{keV}_\text{ee}$ analysis threshold corresponds to 0.84~$\text{keV}_{\text{nr}}$  ($k=0.157$), 1.1 $\text{keV}_{\text{nr}}$ ($k=0.1$), 0.73 $\text{keV}_{\text{nr}}$ ($k=0.2$).  For the power-law model used by CoGeNT, the analysis threshold corresponds to 0.75 $\text{keV}_{\text{nr}}$ . }
		\label{yieldlimit_rito}
	\end{center}
\end{figure}

It is important to understand the systematic effect on our results due to possible inaccuracy in the assumed Lindhard ionization-yield model. The choice of a different yield model systematically changes the nuclear-recoil energy scale, and therefore the interpretation of the data as a limit on the WIMP-nucleon scattering cross section. Figure~\ref{yieldlimit_rito} shows the limits recomputed for four different yield models that bracket the measured data for germanium~\cite{Mei}. A low-ionization Lindhard-like model with $k=0.1$ and a high-yield model with $k=0.2$ are shown, along with the functional form used by the CoGeNT collaboration~\cite{PhysRevD.88.012002}, to demonstrate the effect of this systematic.  The effect of the different yield models is mostly a shift of the limit curve along the WIMP-mass axis. Thus, for masses above $6~\text{GeV}/\text{c}^2$, where the curve is relatively flat, the effect is rather small. For lighter WIMP masses, the systematic uncertainty in yield does produce a noticeable effect on the derived limits.

In conclusion, a very low ionization threshold of $170~\text{eV}_{\text{ee}}$ was achieved with voltage-assisted calorimetric ionization detection, which resulted in sensitivity to light WIMPs. With a small exposure of 6.3~kg-days, and without any background subtraction, new constraints on low-mass WIMPs were obtained. Further exposure will provide more information on the backgrounds, which may allow background subtraction and improve the WIMP sensitivity. The substantial reduction in background levels planned for the SuperCDMS SNOLAB~\cite{Sander:2012nia} experiment would dramatically increase the sensitivity of this experimental mode for low-mass WIMPs. 

The SuperCDMS collaboration gratefully acknowledges  Sten Hansen (PPD, Fermilab), and technical assistance from Jim Beaty and the staff of the Soudan Underground Laboratory and the Minnesota Department of Natural Resources. The iZIP detectors are fabricated in the Stanford Nanofabrication Facility, which is a member of the National Nanofabrication Infrastructure Network sponsored and supported by the National Science Foundation. Part of the research described in this paper was conducted under the Ultra Sensitive Nuclear Measurements Initiative at Pacific Northwest National Laboratory, which is operated by Battelle for the U.S. Department of Energy. Funding and support were received from the National Science Foundation, the Department of Energy, a Fermilab URA Visiting Scholar Award, NSERC Canada and MULTIDARK. Fermilab is operated by the Fermi Research Alliance, LLC under Contract No.~De-AC02-07CH11359. SLAC is operated under Contract No.~DE-AC02-76SF00515 with the United States Department of Energy.

\bibliographystyle{apsrev4-1}


%

\end{document}